\documentclass[fleqn,usenatbib]{mnras}
\usepackage{mathptmx}
\usepackage{amsmath}

\usepackage[T1]{fontenc}
\usepackage{threeparttable}
\usepackage{lscape}
\usepackage{textcomp}
\usepackage{graphicx}
\usepackage{epsfig}
\usepackage{txfonts}
\usepackage{color}
\usepackage{lineno}

\DeclareRobustCommand{\VAN}[3]{#2}
\let\VANthebibliography\thebibliography
\def\thebibliography{\DeclareRobustCommand{\VAN}[3]{##3}\VANthebibliography}

\usepackage{graphicx}	
\usepackage{amsmath}	

\title[Short title, max. 45 characters]{FAST observations of neutral hydrogen in the interacting galaxies NGC 3395/3396}

\author[Nai-Ping Yu et al.]{
Nai-Ping Yu,$^{1,2,3}$\thanks{E-mail: npyu@bao.ac.cn }
Ming Zhu,$^{1,2,3,4}$
Jin-Long Xu$^{1,2,3}$
Chuan-Peng Zhang$^{1,2,3}$
Hai-Yang Yu$^{1,2,4}$
\newauthor
Xiao-Lan Liu$^{1,2,3}$
Peng Jiang$^{1,2,3}$
and Mei Ai$^{1,2,3}$
\\
$^{1}$National Astronomical Observatories, Chinese Academy of Sciences, Beijing 100101, China\\
$^{2}$Guizhou Radio Astronomical Observatory, Guizhou University, Guiyang 550000, People's Republic of China\\
$^{3}$CAS Key Laboratory of FAST, National Astronomical Observatories, Chinese Academy of Sciences, Beijing 100101, China\\
$^{4}$University of Chinese Academy of Sciences, Beijing 100049, China
}

\date{Accepted XXX. Received YYY; in original form ZZZ}

\pubyear{2015}

\begin{document}
\label{firstpage}
\pagerange{\pageref{firstpage}--\pageref{lastpage}}
\maketitle

\begin{abstract}
We report on high-sensitivity neutral hydrogen observations toward the gas-rich interacting galaxies NGC 3395/3396 with the Five-hundred-meter Aperture Spherical radio Telescope (FAST). Compared to previous observations carried out by the Very Large Array (VLA) and the Westerbork Synthesis Radio Telescope (WSRT),  a more extended HI envelope around this system has been detected. The total HI gas mass of the NGC 3395/3396 system is estimated to be 7.8 $\times$ 10$^{9}$ M$_\odot$. This value is 2.7 times more than that reported based on the VLA interferometric maps. Previous observations found a large HI tail extending to the south-west and a minor tail emerging from the north of this peculiar galaxy pair. Based on the high-sensitivity observations of FAST, an extended HI plume to the north-west and a gas plume to the north-east have been detected for the first time. Neutral hydrogen of the two smaller galaxies IC 2604 and IC 2608 on the south of the system have also been detected. We discuss the origins of these extra gas and possible tidal interactions between these galaxies. NGC 3395/3396's most prominent tidal feature, the south-west tail combined with the new detected north-west plume behaves like a large ring. We suggest the ring might be formed by the previous fly-by interaction between NGC 3395 and NGC 3396 which happened 500 Myr ago. Our study shows that high-sensitivity HI observations are important in revealing low column density gas, which is  crucial to a deeper understanding of this interacting system.
\end{abstract}




\section{Introduction}
It is widely accepted that galaxy interactions such as collisions and mergers play an important role in galaxy evolution. These processes can change the morphology of a galaxy and maintain the stable star formation rate inside (e.g. Sancisi et al. 2008; Richter 2017). Atomic hydrogen (HI) is the most abundant component of the interstellar medium (ISM). In gas-rich galaxies, the extent of the HI emission could be several times larger than their optical  diameters ($D_{25}$), making it a very good tracer of tidal interactions, cold gas accretions and/or mergers (e.g. Sancisi et ai. 2008; Serra et al. 2012; Xu et al. 2021). Neutral hydrogen 21cm line surveys of large samples of galaxies indicate that at least one out of four galaxies shows signs of merger or accretion events in the recent past (Sancisi et al. 1999). The interaction history between a galaxy and its environment could be studied by gathering information of extended HI structures such as tails, bridges, plumes and ring-like structures in galaxy surroundings. Many authors have also attempted to explain these complex HI features by means of simple dynamical models (e.g. Toomre $\&$ Toomre 1972; Barnes 1992; Durrell et al. 2003; De Looze et al. 2014). Numerical simulations predict that mergers between two gas-rich disk galaxies could trigger gaseous inflows and starbursts (e.g. Hernquist 1989; Mihos et al. 1991; Mihos $\&$ Hernquist 1996). NGC 3395/3396, also named Arp 270 (Arp 1966), is a pair of closely interacting late-type starburst galaxies. NGC 3395 (the west galaxy) and NGC 3396 (the east galaxy) are classified as Sc and Sm by HyperLeda\footnote{https://leda.univ-lyon1.fr}. The distance to the system is 21.8 Mpc (Tully, Shaya \& Pierce 1992). The optical image shows NGC 3395 is almost face-on, and NGC 3396 is almost edge-on. Their systemic velocities are 1605 and 1678 km s$^{-1}$ respectively according to Garrido et al. (2002), indicating the merger is occurring in a plane almost perpendicular to the line of sight. The system is an approximately equal-mass merger, with a mass ratio of 1.5:1 (Zaragoza-Cardiel et al. 2013). Using the high-resolution $Chandra$ X-ray observation, Brassington et al. (2005) detected sixteen X-ray point sources in the two galaxies, seven of which are classified as ultraluminous X-ray sources. By comparing the luminosities and galaxy properties, they suggest NGC 3395/3396 is in a later stage of evolution than NGC 4485/4490, but earlier than the Antennae and NGC 3256.  Zaragoza-Cardiel et al. (2013) made H$\alpha$ observations of the system. Outflows, characteristic of galactic superwinds and gas inflow toward the centre were detected. They speculate the presence of an active galactic nuclei (AGN) in NGC 3396. Radio continuum observations also found evidences for recent star formation both in the two galaxies (Huang et al. 1994). Using the observations of VLA, Clements et al. (1999) detected a large HI tail extending to a projected distance of 63 kpc south-west of NGC 3395. Besides, a much weaker HI tail emerging from the north of the pair has also been detected (Fig. 1). Smith et al. (2010) discovered a candidate tidal dwarf galaxy at the western extremity of the large tail with moderate optical and UV colors. They also found a faint diffuse emission near the middle of this tail. A comparison with the Mice highlights the absence of any gaseous outflows from the NGC 3395/3396 system (Read 2003).These studies suggest that this interacting galaxy pair is at an earlier stage of evolution than the 11 ongoing mergers listed by Toomre (1977). By means of $N$-body simulations, Clements et al. (1999) conclude that the galaxies are before or after 50 Myr of their second perigalactic passage, with the first approach occurring approximately 500 Myr ago. The current star formation activity is due to the second close approach. They also suggest the NGC 3395/3396 system will merge within 100 Myr. 

Although the NGC 3395/3396 system has been studied in a wide range of wavelengths, we make deeper HI observations of this region with FAST. FAST high-sensitivity observations have revealed many new structures of galaxies in the local universe, such as extra-planar HI clouds and an HI tail in the M101 galaxy group (Xu et al. 2021), a long HI accretion stream toward M106 (Zhu et al. 2021), a 0.6 Mpc HI structure associated with Stephan's Quintet (Xu et al. 2022), and new tidal features in M51 (Yu et al. 2023). In this paper, we report the high-sensitivity HI observations of the interacting galaxies NGC 3395/3396 by FAST. This paper is organized as follows: in Section 2, we introduce our observations and data reduction. Results and analysis are in Section 3, and finally we summarize in Section 4.

\begin{figure}
\center
\psfig{file=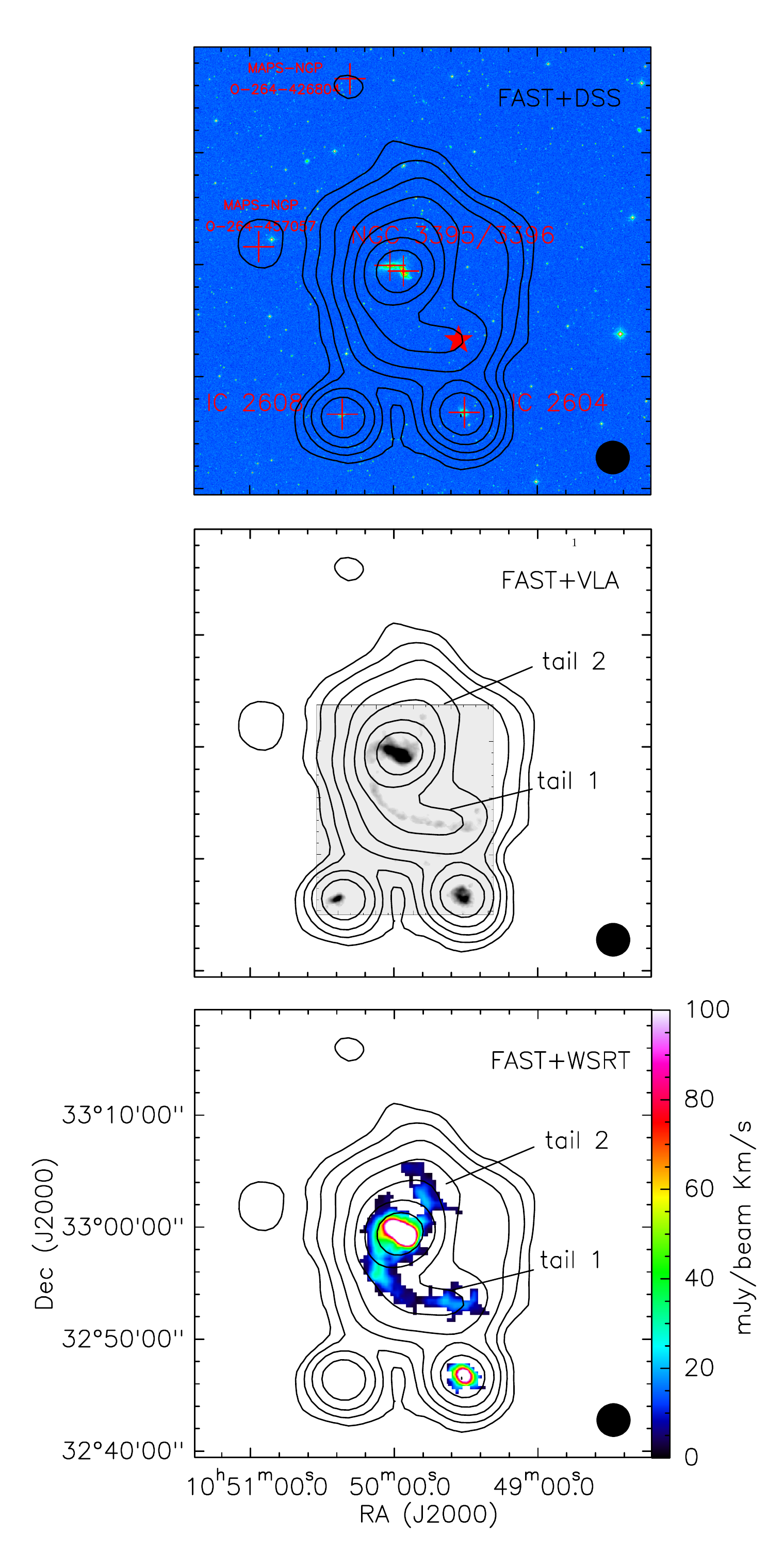,width=3.5in,height=7in}
\caption{Top: The Digitized Sky Survey (DSS) $R$-band optical image of the NGC 3395/3396 region. The black contours are HI column density obtained by FAST, and levels are 2.9 $\times$ 10$^{18}$, 5.9 $\times$ 10$^{18}$, 1.2 $\times$ 10$^{19}$, 2.4$\times$ 10$^{19}$, 4.8 $\times$ 10$^{19}$, 9.6 $\times$ 10$^{19}$ and 19.2 $\times$ 10$^{19}$ cm$^{-2}$. The red crosses mark the center of galaxies and the red star marks the tidal dwarf galaxy candidate found by Clements et al. (1999). Middle: HI column density contours whose levels are consistent with that in the left panel overlaid on the VLA HI image (Clemens et al. 1999). Bottom: HI column density contours whose levels are consistent with that in the left panel overlaid on the WSRT HI image from the WHISP (Westerbork survey of HI in SPiral galaxies) survey (van der Hulst et al. 2001). }
\end{figure}

\section{Observations and data reduction}
We have carried out the FAST All Sky HI Survey (FASHI), which is a new survey of HI emission covering the whole FAST sky between -14$^\circ$ and +66$^\circ$ of declination over the frequency range of 1.0 - 1.5 GHz. The scientific goal of FASHI is to take advantage of the high-sensitivity of FAST to make a complete census of the HI gas in the nearby universe.  The aperture of FAST is 500 m, and its effective aperture is about 300 m. The sensitivity of this telescope reaches 2000 m$^2$/K and the system noise temperature has been controlled below 20 K and the pointing accuracy is $\sim$ 7.9$^{\prime\prime}$ (Jiang et al. 2019, 2020). It is the most sensitive single dish telescope in this frequency range, allowing us to take sensitive observations for various scientific goals, such as studies of pulsars (e.g. Feng et al. 2021; Pan et al. 2021), fast radio burst (e.g. Li 2021), and galaxy evolution (e.g. Zhu et al. 2021; Xu et al. 2022) . From August 2020 to January 2023, more than 7500 square degrees of the sky have been observed, and more than 40000 extragalactic HI galaxies have been detected by FASHI. More information of FASHI is described by Zhang et al. (2023). 

\begin{table*}
\centering
\begin{minipage}{10cm}
\caption{Properties of HI features.}
\begin{threeparttable}
\begin{tabular}{l*{5}c}
      \hline
          Name    &  Velocity range        &  Mass\tnote{a}  &  Mass\tnote{b}  & Mass\tnote{c}  \\
                        &  (km s$^{-1}$)          &    ($\times$10$^8$ M$_\odot$)      &      ($\times$10$^8$ M$_\odot$)       &     ($\times$10$^8$ M$_\odot$)     \\\hline
     Envelope   & 1490-1785   &  29    & 48  &  78   \\
             Tail 1  & 1554-1684   & 4.4     & 7.6  & 18  \\
             Tail 2  & 1660-1692   &   ...     &  2.6 &  4.3 \\
        North-west  Tail     & 1587-1643   & ...    & ... & 4.6 \\
        North-east plume  & 1708-1741   & ...    & ... & 1.1 \\
             IC 2604      &   1542-1718     & 8.5    & 9.0  & 10.2 \\
             IC 2608      &   1583-1773     & 3.5    & ...  & 4.9 \\
          \hline
\end{tabular}
\end{threeparttable}
\begin{tablenotes}
\footnotesize
\item{a} : HI mass calculated from VLA observations with beamsize of $\sim$ 30 arcsec by Clemens et al. (1999).
\item{b} : HI mass calculated from WSRT observations with beamsize of 60 arcsec (van der Hulst et al. 2001).
\item{c} : HI mass calculated from FAST observations with beamsize of 180 arcsec.
\end{tablenotes}
\end{minipage}
\end{table*}

In order to increase the sensitivity and confirm the weak detections, we have further used the Multibeam on-the-fly (OTF) mode to map the NGC 3395/3396 region twice during 2022 January and February. The the 19-beam receiver system in dual polarization mode was used as front-end and the half-power beam width is 2.9 arcmin at 1.4 GHz for each beam. For the digital back-end system, we chose the Spec(W) spectrometer which has a bandwidth of 500 MHz and 65 536 channels, resulting in a velocity resolution of 1.67 km s$^{-1}$ at 1.4 GHz. The data were reduced mainly using the FAST spectral data reduction pipeline HiFAST (Jing et al., 2024; Xu et al., 2023). The HiFAST pipeline combines several useful data reduction packages including antenna temperature correction, baseline correction, RFI mitigation, standing-waves correction, gridding, flux correction, and generating fits cubes. More detailed procedures are described in the HiFAST pipeline cookbook online\footnote{https://hifast.readthedocs.io/}.

\begin{figure*}
\center
\psfig{file=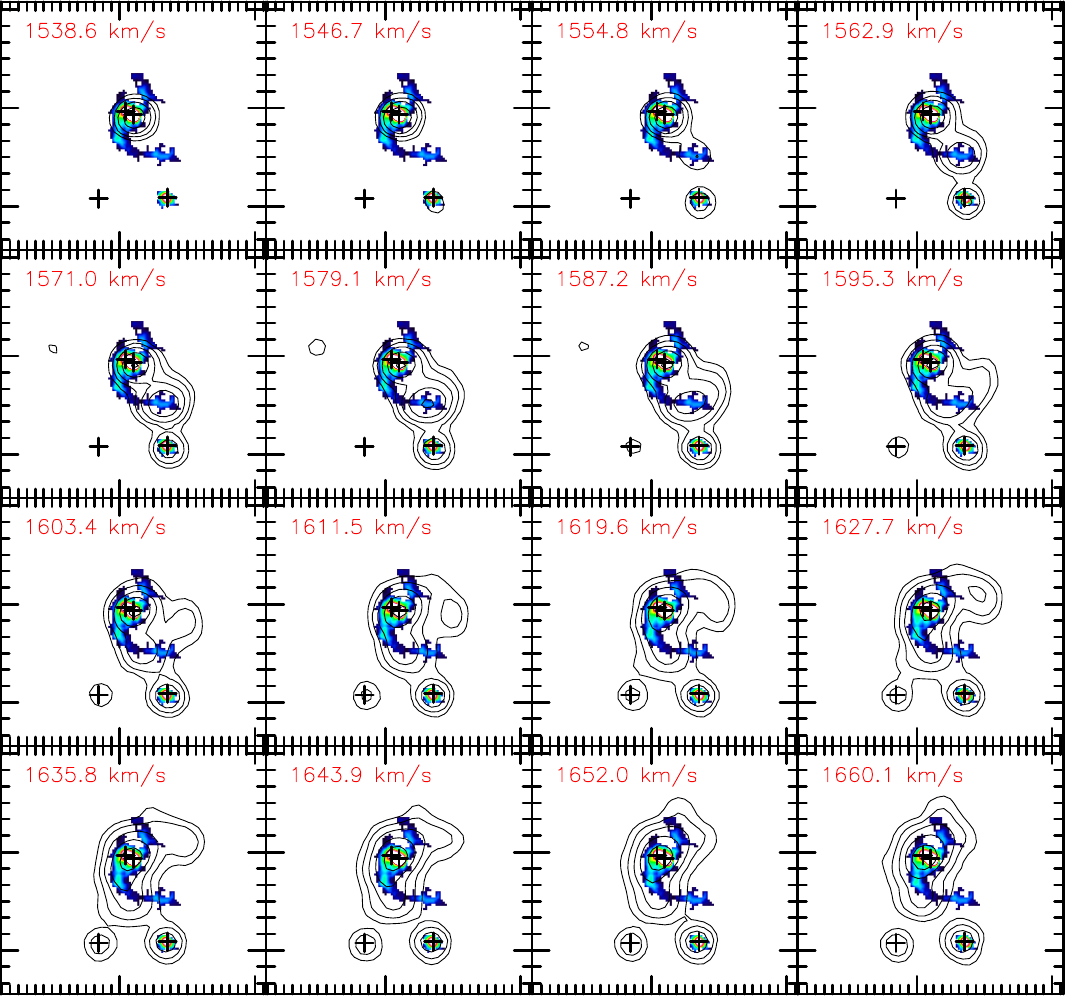,width=5in,height=5in}
\caption{The FAST HI channel maps of this region superimposed on the WSRT moment-0 image. Contour levels are HI column densities of 1.5 $\times$ 10$^{18}$, 3.0 $\times$ 10$^{18}$, 6.0 $\times$ 10$^{18}$, 1.2 $\times$ 10$^{19}$, 2.4 $\times$ 10$^{19}$, 4.8 $\times$ 10$^{19}$ and 9.6 $\times$ 10$^{19}$ cm$^{-2}$. The four black pluses mark the center of NGC 3395, NGC3396, IC 2604 and IC 2608.}
\end{figure*}

\begin{figure*}
    \setcounter{figure}{1}
\center
\psfig{file=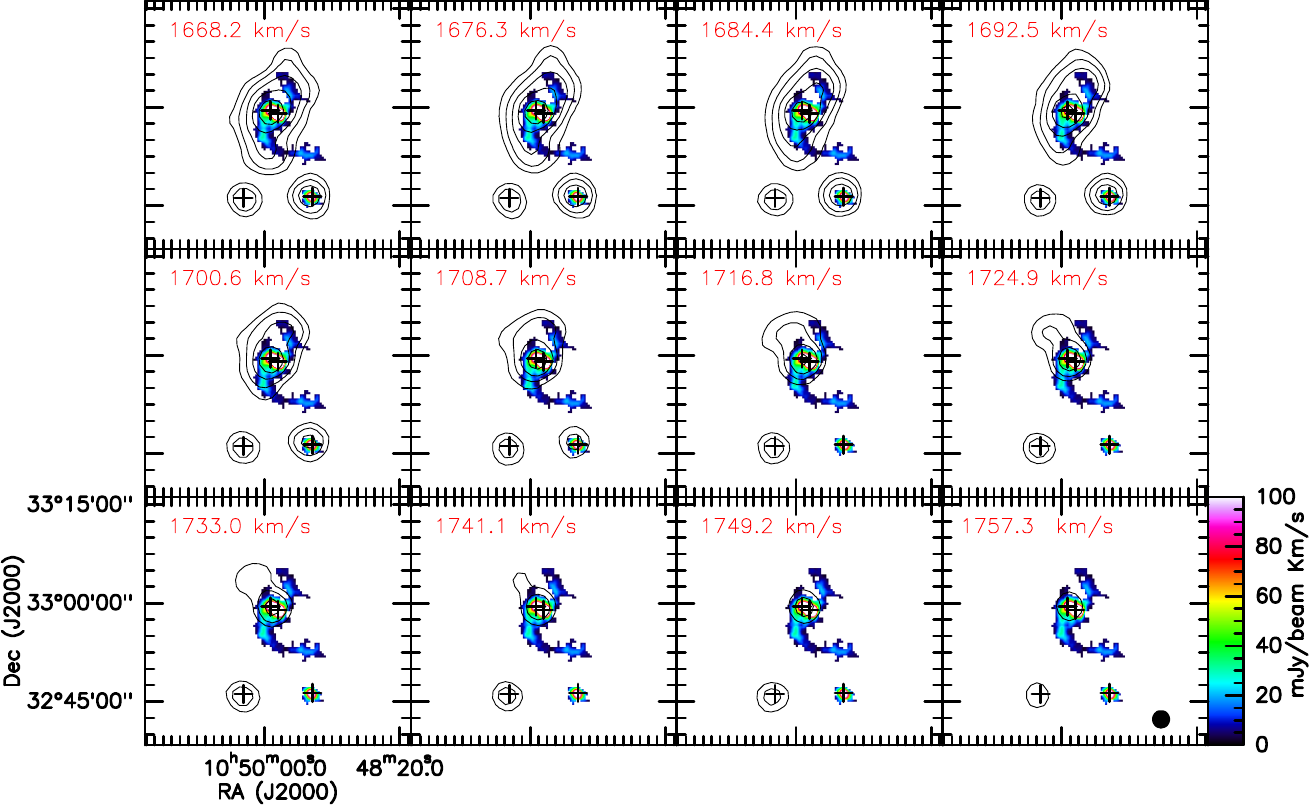,width=6in,height=4in}
\caption{(continued)}
\end{figure*}

\begin{figure*}
\center
\psfig{file=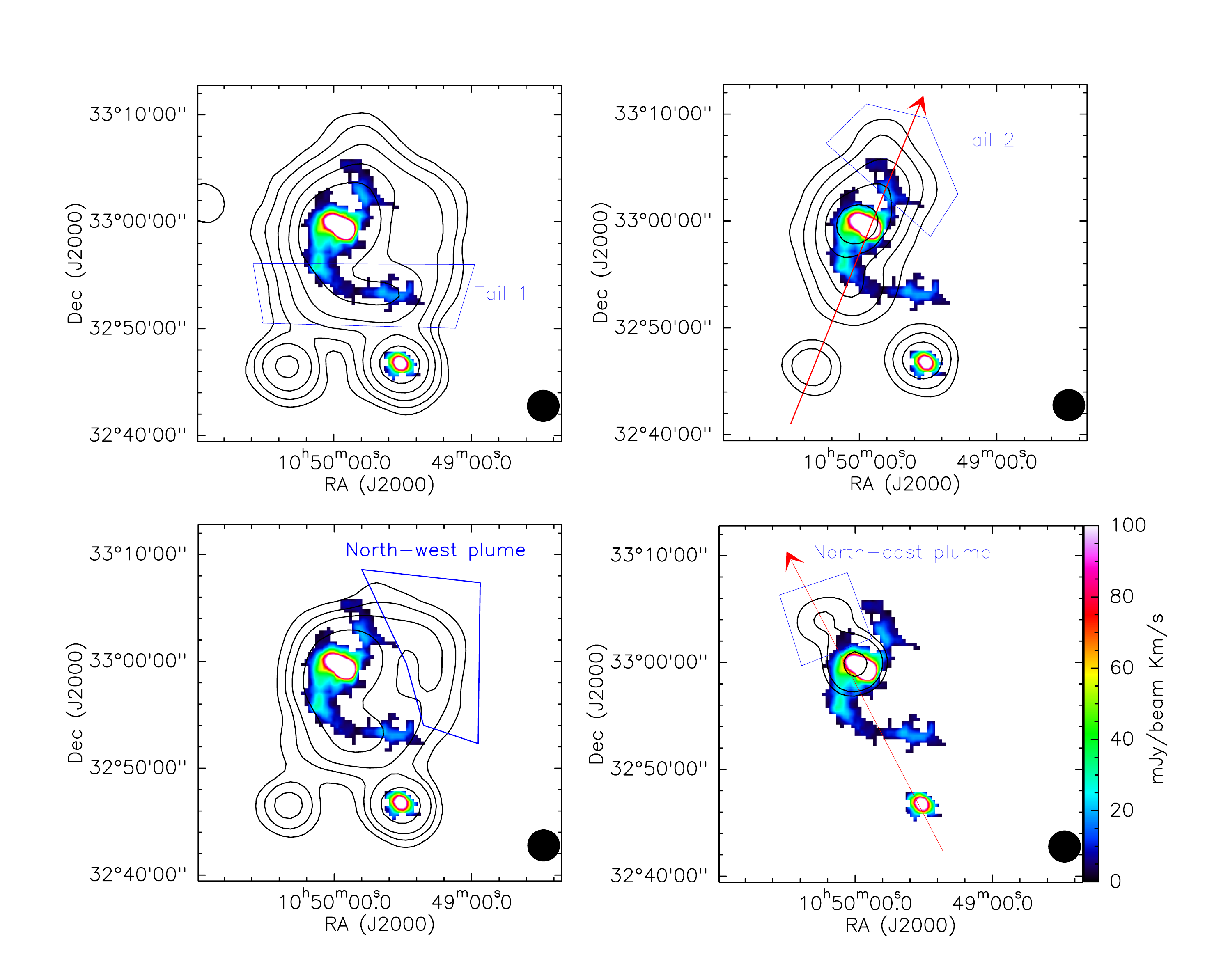,width=5in,height=4in}
\caption{HI column density contours of tail 1 (top left), tail 2 (top right), north-west plume (bottom left) and north-east plume (bottom right) superimposed on the WSRT moment-0 image. Their integrated velocity ranges are listed in Table 1. Contour levels are 2.9 $\times$ 10$^{18}$, 5.9 $\times$ 10$^{18}$, 1.2 $\times$ 10$^{19}$, 2.4 $\times$ 10$^{19}$, and 4.8 $\times$ 10$^{19}$ cm$^{-2}$. The two red arrows show the positions and directions of the PV diagrams shown in Fig.5. }
\end{figure*}

\section{Results and analysis}
An integrated HI flux of 23.2 Jy km s$^{-1}$ is detected in the central region of NGC 3395. This value is consistent with previous observations by Arecibo and the Nan\c{c}ay Radio Telescope (NRT), whose values are 20.2 and 22.0 Jy km s$^{-1}$ respectively (Lewis 1985; van Driel et al. 2001). At a velocity resolution of 6.7 km s$^{-1}$, the noise of our data is 0.5 mJy beam$^{-1}$. At this noise level, the 5$\sigma$ column density threshold within one velocity resolution element is 7.5 $\times$ 10$^{16}$ atoms cm$^{-2}$. This value is about two orders of magnitude lower than that of VLA, which enable us to detect new features in this system. The obtained 21-cm HI image of the NGC 3395/3396 region with FAST, VLA and WSRT are shown in Fig.1. The VLA and WSRT HI images show an extended S-shaped structure around the pair, with a large tail in the south-west (we call it ``tail 1" hereafter in this paper) and a minor tail (hereafter ``tail 2") emerging from the north of the system. The nearby company galaxies IC 2604 and/or IC 2608 are also detected by VLA and WSRT. It can be noticed that the HI emission detected by FAST is more extended. The HI gas of IC 2604 and IC 2608 seems to be connected with the envelope of NGC 3395/3396. The most prominent detected HI features are the gas in the north-west and north-east of NGC 3395/3396. These structures have been detected for the first time. The discussions of these structures are in the following subsection 3.2. The total gas mass of the envelope of NGC 3395/3396 could be estimated by equation
\begin{equation}
M = 2.35 m D^2 \int N(x,y) dxdy
\end{equation}
where the factor 2.35 is the mean atomic weight, $m$ is the mass of atomic hydrogen, $D$ is the distance, $dx$ and $dy$ are the pixel sizes, $N(x,y)$ is the HI column density which could be calculated through
\begin{equation}
N (x,y) = 1.1 \times 10^{24} \frac{\int S_v dv}{\theta_{beam}^2}
\end{equation}
where $S_v$ is the flux density in unit of Jy, and $\theta$ is the beam size of FAST in arcsec. The obtained gas mass of the envelope is 7.8 $\times$ 10$^9$ M$_\odot$, about 3 times higher than that found by Clements et al. (1999), indicating more diffuse cold gas has been detected.

\subsection{The south-east tail and north tail}
Previous HI observations have already found two HI tails in the gas-rich interacting system of NGC 3395/3396. Fig.2 shows our FAST HI channel maps of this region superimposed on the WSRT moment-0 image. We can see that tail 1 has an HI velocity range of 1554.8 km s$^{-1}$ to 1684.4 km s$^{-1}$, and tail 2 has a velocity range of 1660.1 km s$^{-1}$ to 1692.5 km s$^{-1}$. This is consistent with the observations of VLA (Clements et al. 1999) and WSRT (van der Hulst et al. 2001). Clemens et al. (1999) first discovered these tails. According to their study, tail 1 has been stripped from NGC 3395 during a prograde encounter with NGC 3395, although it appears to be joined with NGC 3396 in the VLA map. By means of $N$-body simulations, they further found NGC 3395/3396 is within 50 Myr of its second close approach, which may have just occurred or will occur in the near future. The FAST integrated HI emission of the south-east tail is shown in the top left panel of Fig.3. The presence of tail 1 is wider and longer than those revealed by VLA and WSRT, indicating the existence of more diffuse and faint gas. Using the above two equations and the integrated velocity range shown in table 1, we estimate the HI gas mass of tail 1 to be 1.8 $\times$ 10$^9$ M$_\odot$, which is 4.1 and 2.3 times higher than those detected by VLA and WSRT, respectively. The mass of tail 1 is 23 percent of the total envelope of NGC 3395/3396. This value is much lower than that found in the tails of the Antennae, where 70 percent is found (van der Hulst 1979).

Tail 2 is a much weaker HI tail emerging from the north of the NGC 3305/3396 system (Clemens et al. 1999). This tail is more obvious in the HI image of WSRT (Fig.1). In our observations, this tail is hardly resolved owing to the large beam of FAST. By comparison with the HI and optical images, Clemens et al. (1999) suggest that this minor tail is the extension of an HI ridge which coincides with an optical tidal feature in the south of NGC 3395. This tidal feature could also be reproduced in the simulations of Clemens et al. (1999), but needs slight changes to increase the projected length.

\subsection {The north-west and north-east plumes}
\begin{figure}
\center
\psfig{file=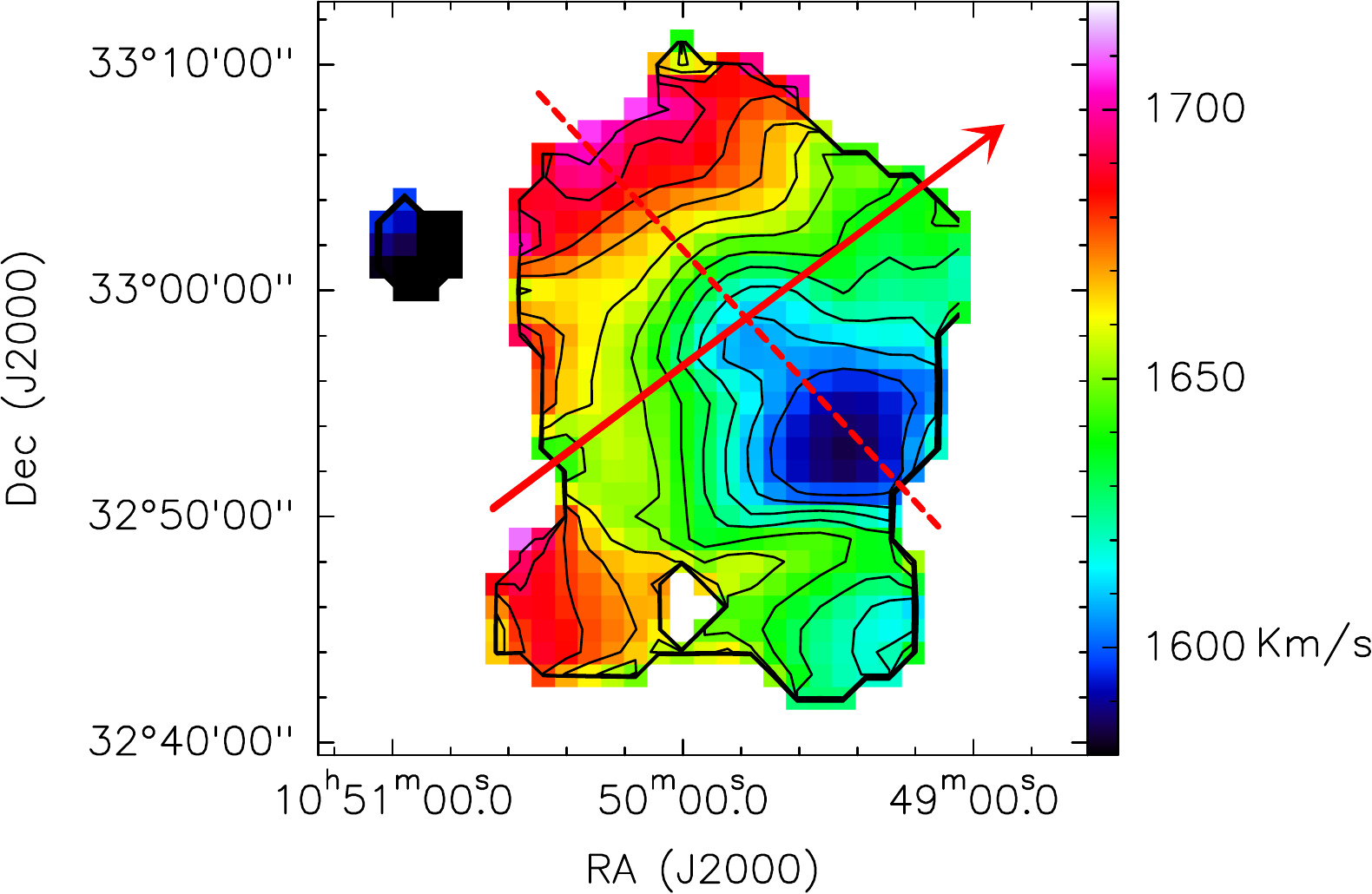,width=3in,height=1.8in}
\psfig{file=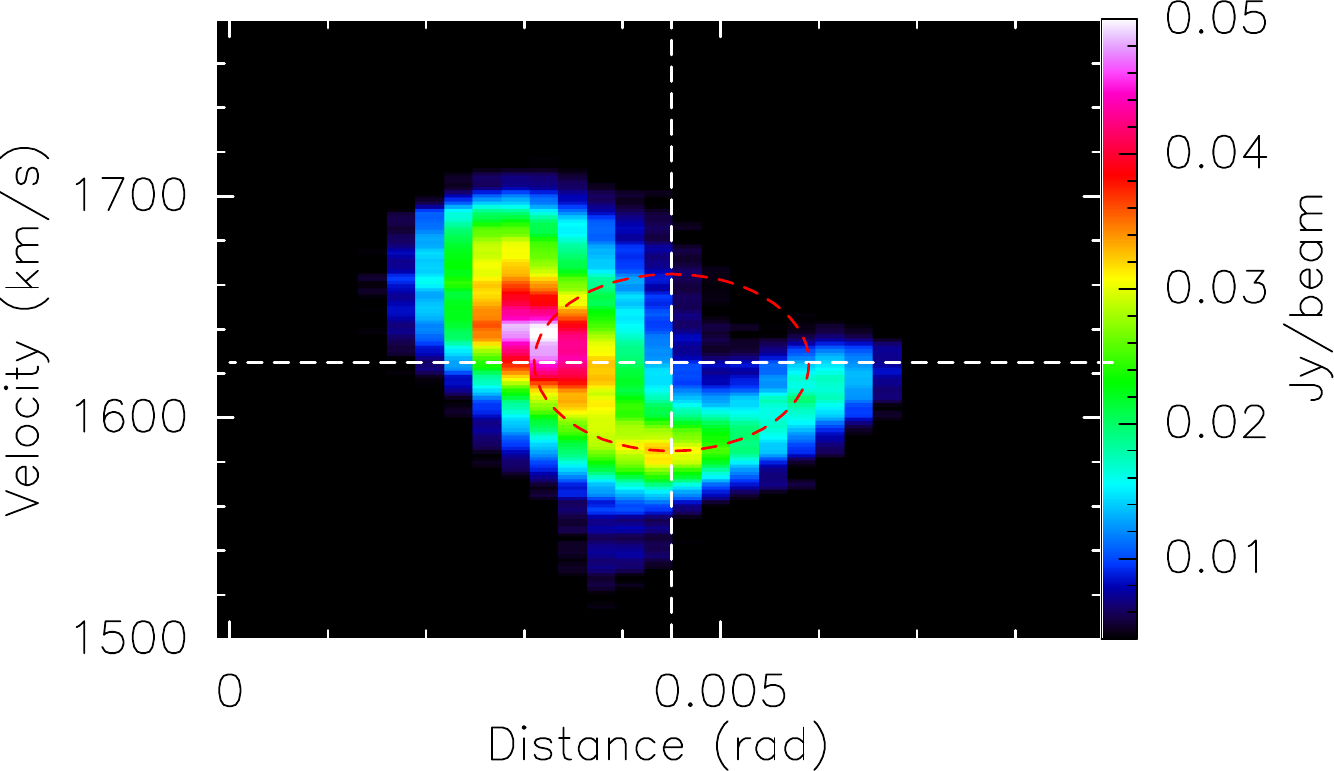,width=3.2in,height=1.8in}
\caption{Top: Moment-1 map (velocity field) of the region with contours and colourscale. Contour levels go from 1600 to 1700 km s$^{-1}$ in step of 10 km s$^{-1}$. The red arrow indicates the position and direction of the PV diagram shown in the bottom panel, and the dashed red line shows the symmetric axis of the velocity field. Bottom: PV diagram of HI data cube cut through the ring. The red dashed ellipse shows a shell with radius of 29.8 kpc and expanding velocity of 40 km s$^{-1}$, and the two white dashed lines show the center and systemic velocity of the ring respectively.  }
\end{figure}

These two HI features are detected by FAST for the first time. In the channel maps of Fig.2, we notice that from the velocity range of 1587.2 km s$^{-1}$ to 1643.9 km s$^{-1}$, there is an extra plume of HI gas to the north-west of NGC 3395/3395. This plume seems to stem from the western extremity of tail 1 and connects to the north-west of the envelope of NGC 3395/3396. The bottom left panel of Fig.3 displays the integrated flux density map of this extra plume. It looks like a tail on the opposite side of tail 1. Fig.4 is the moment-1 map (velocity field) of the NGC3395/3396 system. The envelope of NGC 3395/3396 has a velocity gradient from north-east to south-west. The velocity fields of tail 1 and the north-west plume are almost symmetrical, with a symmetric axis from northeast to southwest, suggesting the two tails might be two halves of a large gas ring. Ring like structures are common around interacting galaxies (e.g. Schneider et al. 1989; Malphrus et al. 1997; Ryder et al. 2001). According to the models of Toomre (1978), ring morphology could be produced by an off-center collision between two galaxies, and the far side of the ring will soon open up into extended tidal tails. The $N$-body simulation of Clemens et al (1999) suggests that a weak fly-by interaction between NGC 3395 and NGC 3396 has taken place about 500 Myr ago. The bottom panel of Fig.4 shows the position–velocity (PV) diagram cut through the center of the ring. The velocity in the center of the ring is lowest ($\sim$ 1587 km s$^{-1}$), and increases towards the two tails, indicating the ring is expanding. The expanding velocity ($v_{exp}$) of the ring could be estimated by measuring the deviation of the velocity of the ring seen in the PV diagram (Pokhrel et al. 2020). We calculate $v_{exp}$ as the difference between the systemic velocity of the ring and the gas velocity at the center (where the gas is deviating from the surroundings). We adopt 1627 $\pm$ 8 km s$^{-1}$ as the systemic velocity of the ring as the ring feature is clearest in the velocity range between 1619.6 and 1635.8 km s$^{-1}$ in the channel maps of Fig. 2. The expanding velocity of the ring is thus estimated to be about 40 $\pm$ 8 km s$^{-1}$. The expansion age of the ring could thus be estimated by equation $t_{exp}$ = $R$/$v_{exp}$, where $R = 29.8 kpc$ is the radius of the ring. The derived value is 725 $\pm$ 145 Myr, which is roughly consistent with the simulation of Clemens et al (1999). We thus speculate that tail 1 and the north-west plume are two half parts of a ring which was produced by the first encounter of NGC 3395 and NGC 3396 about 500 Myr ago. 

From the channel maps of Fig.2, we found another new structure on the north-east of NGC 3395/3396 in the velocity range of  1708.7 km s$^{-1}$ to 1741.1 km s$^{-1}$. Its integrated flux density is also shown in the bottom right panel of Fig.3. The HI gas mass is 1.1 $\times$ 10$^8$ M$_\odot$, which is the lowest among the four structures. We check the optical images from Sloan Digital Sky Survey (SDSS)\footnote{https://dr12.sdss.org/fields}, the Dark Energy Spectroscopic Instrument (DESI) legacy imaging surveys\footnote{https://www.legacysurvey.org/decamls/} and ultraviolent images from the Galaxy Evolution Explore (GALEX)\footnote{https://galex.stsci.edu/GR6/},  which are all publicly available on line. Emission in these wavelengths by the north-east plume is extremely deficient. No previously catalogued galaxy matches with this HI feature. A search in this region on the NED/IPAC extragalactic database (NED) also yields no findings.

\subsection{Possible tidal interactions with IC 2604 and IC 2608}
\begin{figure}
\center
\psfig{file=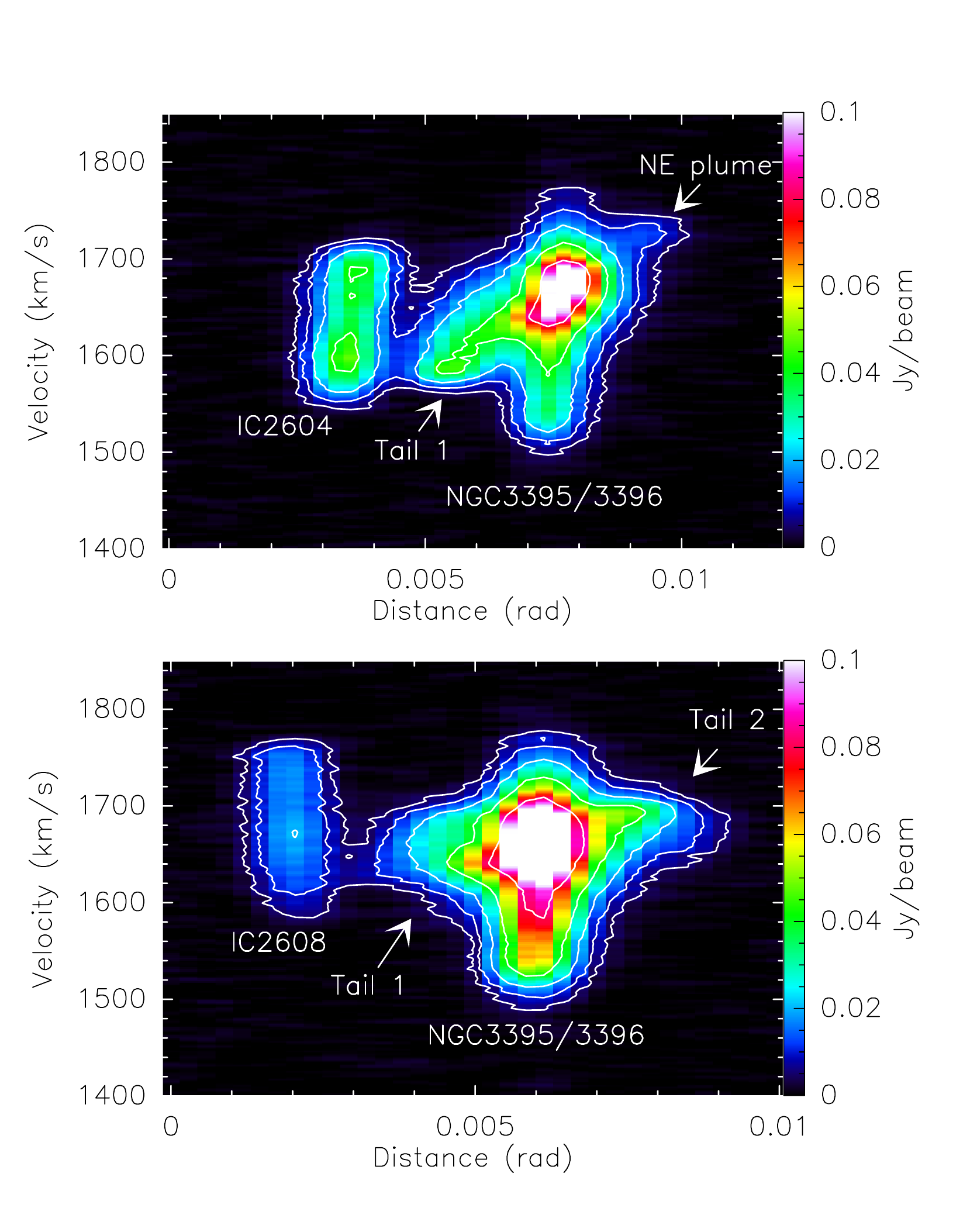,width=3in,height=4in}
\caption{The PV diagrams of HI data cube cut along directions and positions shown in Fig. 3. Contour levels are 5, 10, 20, 40, 80 and 160 mJy/beam.}
\end{figure}

\begin{figure*}
\center
\psfig{file=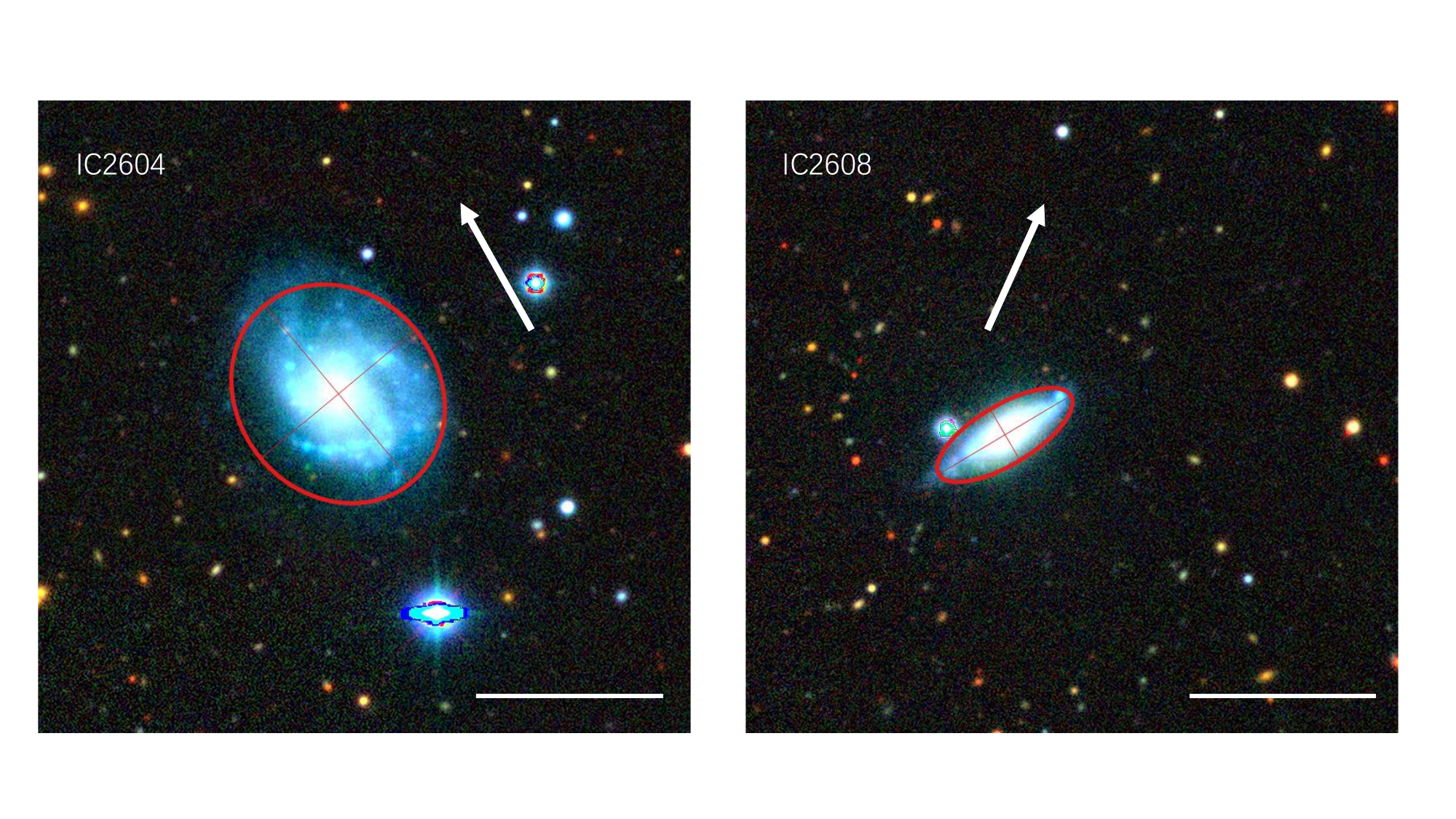,width=5in,height=3.0in}
\caption{Composite color images of IC 2604 (left) and IC 2608 (right): g band in blue, r band in green, and z band in red. The red ellipses show their optical disks from the HyperLeda data base (Makarov et al. 2014). The white arrow shows the direction to NGC 3395/3396. The white line on the bottom right indicates 5 kcp along the horizontal axis.}
\end{figure*}

 The HI emission of the two smaller galaxies IC 2604 which is 14 arcmin to the south-west, and IC 2608 which is 14 arcmin to the south-east have also been detected by VLA and FAST (Fig.1). The HI emission of IC 2608 is undetected by WSRT. The HI velocities of the two galaxies range from 1546 km s$^{-1}$ to 1708 km s$^{-1}$ and 1587 km s$^{-1}$ to 1757 km s$^{-1}$, respectively, indicating they lie at the similar distance as the NGC 3395/3396 system. IC 2604 is classified as a barred spiral by de Vaucouleurs et al. (1991). The optical image of this galaxy appears asymmetrical with a large HII region at the extremity of its southern arm. Its HI morphology is also asymmetrical with the peak emission in this optical arm (Clemens et al. 1999). IC 2608 is an Sc-type galaxy (Vaucouleurs et al. 1991). The optical disk of this galaxy appears almost edge-on. The HI peak emission is coincident with the optical center, with a small tail extending to the south-east (see the Fig.2 of Clemens et al. 1999). The asymmetrical HI structures of IC 2604 and IC 2608 indicate possible tidal interactions with the NGC 3395/3396 system. From Fig.1, we notice the HI emissions of IC 2604 and IC 2608 detected by FAST are more extended, and seem to be connected with the envelope of NGC 3395/3396. From Table 1, it could also be noticed that the gas mass of these galaxies calculated from the observations of FAST are 20 percent and 42 percent more than those detected by VLA, indicating more diffuse gas has been detected. IC 2604 and IC 2608 may be physically connected to the envelope of NGC 3395/3396 by neutral hydrogen. Clemens et al. (1999) suggested IC 2604 and IC 2608 have at most a minor influence on the dynamics of NGC 3395/3396, as it is not necessary to include them in their $N$-body simulations to reproduce the HI morphology of tail 1 and tail 2. As is shown above, our FAST observations have detected more diffuse gas and new structures in the envelope of NGC 3395/3396, indicating slight changes of their models are needed. 

We now discuss the possible tidal interactions between NGC 3395/3396 and IC 2604/2608. The top panel of Fig.5 shows the PV diagram cut through IC 2604, tail 1, NGC3395/3396 and the north-east plume. The positions and directions of the two PV diagrams are shown as the two red arrows in Fig.3.  It can be noticed the IC 2604 contours are asymmetrical with an HI bridge connecting to tail 1 in the PV diagram, implying they are interacting. The north-east plume might be a tidal tail on the opposite side of NGC 3395/3396 caused by IC 2604. Fig.6 shows the composite color images of IC 2604 and IC 2608 from the DESI legacy imaging surveys (Dey et al. 2019). The star distribution of IC 2604 is asymmetrical and seems more extended to the north-east, which is the direction to the NGC 3395/3396 system. Our study suggests tidal interaction with IC 2604 should be considered in the study of the evolution of the NGC 3395/3396 system. More gas is stripped toward tail 1 by the tidal force of IC 2604, making this tail  much denser and more massive than the north-west plume. This may be the reason the mass of the north-west plume is only a quarter of tail 1. The bottom panel of Fig.5 shows the PV diagram cut through IC 2608, tail 1, NGC3395/3396 and tail 2. The IC 2608 contours look symmetrical, even though there is also a weak HI bridge connecting to tail 1. The deep optical image of IC 2608 (Fig. 6) is also symmetrical. Consistent with the study of Clemens et al. (1999), we regard tidal interaction between NGC 3395/3396 and IC 2608 is weak, although 40 percent more diffuse gas has been detected in this galaxy.

In gas-rich interacting galaxies, material could be pulled from the system and forms gaseous filaments that can harbour second generation tidal knots and tidal dwarf galaxies (TDGs).
Clements et al. (1999) noticed a distinct clump of HI in the western extremity of tail 1 (Fig. 1). Dynamical analysis indicates the clump is in virial equilibrium and unable to collapse to form stars. They thus suggest this clump represents another example of TDG. Based on the images from GALEX and SDSS, Smith et al. (2010) found the source near the end of tail 1 has moderate optical and UV colors, indicating star formation activities inside tail 1. Dynamical simulations suggest tidal interactions could trigger starburst by compressing the gas of a galaxy (e.g. Linden $\&$ Mihos 2022).  To quantify the effect of tidal interactions with this TDG, we calculate the tidal index $\Theta$ defined by Karachentsev $\&$ Makarov (1998) as
 \begin{equation}
\Theta = log_{10} (\frac{M_{\ast} [\times10^{11}M_{\odot}]}{D_{project}^3 [Mpc]})
\end{equation}
where $M_{\ast}$ is the stellar mass of the NGC 3395/3396 system (2.4 $\times$ 10$^{10}$ M$_\odot$ according to Zaritsky et al. 2014), IC 2604 (6.3 $\times$ 10$^{8}$ M$_\odot$ according to Leroy et al. 2019) and IC 2608 (4.7 $\times$ 10$^{8}$ M$_\odot$ according to Leroy et al. 2019) respectively. $D_{project}$ is the projected distance from the TDG to these galaxies. A higher $\Theta$ indicates a stronger influence. The derived values of $\Theta$ are 3.2, 2.0 and 1.3 for the NGC 3395/3396 system, IC 2604 and IC 2608, respectively. Pearson et al. (2016) found if a dwarf galaxy pair has a tidal indices $\Theta$ $>$ 1.5, the pair is most likely affected by tidal and/or ram-pressure stripping from the host galaxy. Our study suggests the TDG in tail 1 is affected by tidal interactions mainly from the NGC 3395/3396 system, but the influence from IC 2604 should not be ignored. IC 2608 has at most a minor influence on the dynamics of the NGC 3395/3396 system. Simulations are beyond the scope of this paper. But a new model involving three galaxies (NGC 3395, NGC 3396 and IC 2604) is required to obtain a fully matched result in the future. 
 
\section{Summary}
We report the high-sensitivity HI observations toward the gas-rich interacting galaxies NGC 3395/3396 with FAST.  More diffuse cold gas and new structures around this system have been detected. The total HI gas mass of the envelope of NGC 3395/3396 is 2.7 times more than that reported previously. An extended HI plume on the north-west and a new gas plume on the north-east have been detected for the first time. Neutral hydrogen of the two smaller galaxies IC 2604 and IC 2608 on the south have also been detected. We discuss the origins of these extra gas and possible tidal interactions between these galaxies. NGC 3395/3396's most prominent tidal feature, the south-west tail combined with the north-west plume look like a large ring. The ring might be formed by the previous fly-by interaction between NGC 3395 and NGC 3396. IC 2604 probably has tidal interactions with the NGC 3395/3396 system. Our study shows that high-sensitivity HI observations are crucial in revealing low column density features in nearby galaxies.

\section{Acknowledgements}
We thank the anonymous referee and scientific editor for insightful comments and constructive suggestions. We acknowledge supports from the National Key R$\&$D Program of China No. 2018YFE0202900. We thank Clemens M.S. for sharing their VLA observation on the web. This work was supported by the Youth Innovation Promotion Association of Chinese Academy of Science (CAS), the National Natural Foundation of China No. 12373001, and also supported by the Open Project Program of the Key Laboratory of FAST, NAOC, Chinese Academy of Sciences.

\section{Data availability}
The data underlying this article will be shared on reasonable request to the corresponding author.

\bibliographystyle{mnras}
\bibliography{example} 

\begin{thebibliography}{}
\bibitem[\protect\citeauthoryear{Contreras et al.}{2007a}]{b21} Arp, H., 1966, ApJS, 14,1

\bibitem[\protect\citeauthoryear{Contreras et al.}{2007a}]{b21} Barnes, J., 1992, ApJ, 393, 484

\bibitem[\protect\citeauthoryear{Contreras et al.}{2007a}]{b21} Barnes, J. E., Hernquist, L., 1992, Nature, 360, 715

\bibitem[\protect\citeauthoryear{Contreras et al.}{2007a}]{b21} Brassington, N. J., Read, A. M., \& Ponman, T. J. 2005, MNRAS, 360, 801

\bibitem[\protect\citeauthoryear{Contreras et al.}{2007a}]{b21} Clemens, M. S., Baxter, K. M., Alexander, P., Green, D. A., 1999, MNRAS, 308, 364

\bibitem[\protect\citeauthoryear{Contreras et al.}{2007a}]{b21} de Vaucouleurs G., de Vaucouleurs A., Corwin, H. G., Buta, R. J., Fouque, P., Paturel, G., 1991, Third Reference Catalogue of Bright Galaxies. Springer, New York

\bibitem[\protect\citeauthoryear{Contreras et al.}{2007a}]{b21} De Looze I. et al., 2014, A\&A, 571, A69

\bibitem[\protect\citeauthoryear{Contreras et al.}{2007a}]{b21} Durrell, P. R., Mihos, J. C., Feldmeier, J. J., Jacoby, G. H., Ciardullo, R., 2003, ApJ, 582, 170

\bibitem[\protect\citeauthoryear{Contreras et al.}{2007a}]{b21} Garrido, O., Marcelin, M., Amram, P., Boulesteix, J., 2002, A\&A, 387, 821

\bibitem[\protect\citeauthoryear{Contreras et al.}{2007a}]{b21} Hernquist, L. 1989, Natur, 340, 687

\bibitem[\protect\citeauthoryear{Contreras et al.}{2007a}]{b21} Huang, Z.-P., Yin, Q.-F., Saslaw, W. C., Heeschen, D. S., 1994, ApJ, 423, 614

\bibitem[\protect\citeauthoryear{Contreras et al.}{2007a}]{b21} Jing, Y., Wang, J., Xu, C., et al. 2024, Science China Physics, Mechanics, and Astronomy, 67, 259514 2

\bibitem[\protect\citeauthoryear{Contreras et al.}{2007a}]{b21} Karachentsev, I. D., Makarov, D. I., 1998, A\&A, 331, 891

\bibitem[\protect\citeauthoryear{Contreras et al.}{2007a}]{b21} Leroy, A. K., Sandstrom, K. M., Lang, D., et al. 2019, ApJS, 244, 24

\bibitem[\protect\citeauthoryear{Contreras et al.}{2007a}]{b21} Lewis, B. M. 1985, ApJ, 292, 451

\bibitem[\protect\citeauthoryear{Contreras et al.}{2007a}]{b21} Linden, S. T., Mihos, J. C., 2022, ApJL, 933, L33

\bibitem[\protect\citeauthoryear{Contreras et al.}{2007a}]{b21} Malphrus, B.K., et al. 1997, AJ, 114, 1427

\bibitem[\protect\citeauthoryear{Contreras et al.}{2007a}]{b21} Makarov, D., Prugniel, P., Terekhova, N., Courtois, H., Vauglin, I., 2014, A\&A, 570, A13

\bibitem[\protect\citeauthoryear{Contreras et al.}{2007a}]{b21} Michel-Dansac, L., Duc, P.-A., Bournaud, F., et al. 2010, ApJ, 717, L143

\bibitem[\protect\citeauthoryear{Contreras et al.}{2007a}]{b21} Mihos, J. C., Richstone, D. O., Bothun, G. D., 1991, ApJ, 377, 72

\bibitem[\protect\citeauthoryear{Contreras et al.}{2007a}]{b21} Mihos, J. C., \& Hernquist, L. 1996, ApJ, 464, 641

\bibitem[\protect\citeauthoryear{Contreras et al.}{2007a}]{b21} Pearson, S., Besla, G., Putman, M. E., et al. 2016, MNRAS, 459, 1827

\bibitem[\protect\citeauthoryear{Contreras et al.}{2007a}]{b21} Pokhrel, N. R., Simpson, C. E., Bagetakos, I., 2020, AJ, 160, 66

\bibitem[\protect\citeauthoryear{Contreras et al.}{2007a}]{b21} Read, A. M., 2003, MNRAS, 342, 715

\bibitem[\protect\citeauthoryear{Contreras et al.}{2007a}]{b21} Richter, P., 2017, A\&A, 607, A48

\bibitem[\protect\citeauthoryear{Contreras et al.}{2007a}]{b21} Ryder, S.D., et al. 2001, ApJ, 555, 232

\bibitem[\protect\citeauthoryear{Contreras et al.}{2007a}]{b21} Serra, P., Oosterloo, T., Morganti, R. et al. 2012, MNRAS, 422, 1835

\bibitem[\protect\citeauthoryear{Contreras et al.}{2007a}]{b21} Sancisi, R. 1999, in Galaxy Interactions at Low and High Redshift, eds. J. E.
Barnes, \& D. B. Sanders, IAU Symp., 186, 71

\bibitem[\protect\citeauthoryear{Contreras et al.}{2007a}]{b21} Sancisi, R., Fraternali, F., Oosterloo, T., \& van der Hulst, T., 2008, A\&ARv, 15, 189

\bibitem[\protect\citeauthoryear{Contreras et al.}{2007a}]{b21} Schneider, S.E., et al. 1989, AJ, 97, 666

\bibitem[\protect\citeauthoryear{Contreras et al.}{2007a}]{b21} Toomre, A., Toomre, J., 1972, ApJ, 178, 623

\bibitem[\protect\citeauthoryear{Contreras et al.}{2007a}]{b21} Toomre, A., 1977, in Tinsky, B. M., Larson, T. B., eds, Evolution of Galaxies and Stellar Popultations. Yale Univ. Observatory, NewHaven, p. 401

\bibitem[\protect\citeauthoryear{Contreras et al.}{2007a}]{b21} Toomre, A. 1978, IAU Symp. No. 79, Large Scale Structures in the Universe, Vol. 79 ed. M. S. Longair \& J. Einasto (Dordrecht: Reidel), 109

\bibitem[\protect\citeauthoryear{Contreras et al.}{2007a}]{b21} Theys, J. C., \& Spiegel, E. A. 1977, 212, 616

\bibitem[\protect\citeauthoryear{Contreras et al.}{2007a}]{b21} Tully, R. B., Shaya, E. J., Pierce, M. J., 1992, ApJS, 80, 479

\bibitem[\protect\citeauthoryear{Contreras et al.}{2007a}]{b21} van Driel, W., Marcum, P., Gallagher, J. S., I., et al. 2001, A\&A, 378, 370

\bibitem[\protect\citeauthoryear{Contreras et al.}{2007a}]{b21} van der Hulst J. M., 1979, A\&A, 71, 131

\bibitem[\protect\citeauthoryear{Contreras et al.}{2007a}]{b21} van der Hulst J. M., van Albada T. S., Sancisi R., 2001, in Hibbard J. E., Rupen M., van Gorkom J. H.eds, ASP Conf. Ser. Vol. 240, Gas and Galaxy Evolution. Astron. Soc. Pac., San Francisco. p. 451

\bibitem[\protect\citeauthoryear{Contreras et al.}{2007a}]{b21} Xu, C. K. et al., 2022, Nature, 610, 461

\bibitem[\protect\citeauthoryear{Contreras et al.}{2007a}]{b21} Xu, C., Jing, Y., Wang, J., et al. 2023, RAA, prep, prep

\bibitem[\protect\citeauthoryear{Contreras et al.}{2007a}]{b21} Xu, J.-L. et al., 2021, ApJ, 922, 53

\bibitem[\protect\citeauthoryear{Contreras et al.}{2007a}]{b21} Yu, H. Y., Zhu, M., Xu, J.-L., Ai, M., Jiang, P., Yang, Y. B., 2023, MNRAS, 521, 2719

\bibitem[\protect\citeauthoryear{Contreras et al.}{2007a}]{b21} Zaragoza-Cardiel, J., Font-Serra, J., Beckman, J. E., et al. 2013, MNRAS, 432, 998

\bibitem[\protect\citeauthoryear{Contreras et al.}{2007a}]{b21} Zaritsky, D., Courtois, H., Mu$\tilde{n}$oz-Mateos, J.-C., et al. 2014, AJ, 147, 134

\bibitem[\protect\citeauthoryear{Contreras et al.}{2007a}]{b21} Zhang, C. P., Zhu, M., Jiang, P. et al. 2024, SCPMA, 67, 219511 

\bibitem[\protect\citeauthoryear{Contreras et al.}{2007a}]{b21} Zhu, M. et al., 2021, ApJ, 922, L21

\bibitem[\protect\citeauthoryear{Contreras et al.}{2007a}]{b21} Zwicky, F., 1956, Ergeb. Exact. Nat., 26, 344

\end{thebibliography}


\bsp	
\label{lastpage}
\end{document}